\def\presentation{
\voffset -.50in  \hoffset -.19in
\oddsidemargin 0in \evensidemargin 0in
\marginparwidth .75in \marginparsep 7pt \topmargin 0in
\headheight 12pt \headsep .25in
\footheight 18pt \footskip .35in
\textheight 9.5in \textwidth 6.5in
\columnsep 10pt \columnseprule 0pt }
\documentstyle{article}
\presentation
\begin{document}
%

%
\def\tilde{\widetilde}
\def\bar{\overline}
\def\hat{\widehat}
\def\*{\star}
\def\[{\left[}
\def\]{\right]}
\def\({\left(}
\def\){\right)}
\def\zb{{\bar{z} }}
\def\frac#1#2{{#1 \over #2}}
\def\inv#1{{1 \over #1}}
\def\half{{1 \over 2}}
\def\d{\partial}
\def\der#1{{\partial \over \partial #1}}
\def\dd#1#2{{\partial #1 \over \partial #2}}
\def\vev#1{\langle #1 \rangle}
\def\bra#1{{\langle #1 |  }}
\def\ket#1{ | #1 \rangle}
\def\rvac{\hbox{$\vert 0\rangle$}}
\def\lvac{\hbox{$\langle 0 \vert $}}
\def\2pi{\hbox{$2\pi i$}}
\def\e#1{{\rm e}^{^{\textstyle #1}}}
\def\grad#1{\,\nabla\!_{{#1}}\,}
\def\dsl{\raise.15ex\hbox{/}\kern-.57em\partial}
\def\Dsl{\,\raise.15ex\hbox{/}\mkern-.13.5mu D}
\def\comm#1#2{ \BBL\ #1\ ,\ #2 \BBR }
\def\x{\stackrel{\otimes}{,}}
\def\det{ {\rm det}}
\def\tr{{\rm tr}}
%
%
\def\th{\theta}		\def\Th{\Theta}
\def\ga{\gamma}		\def\Ga{\Gamma}
\def\be{\beta}
\def\al{\alpha}
\def\ep{\epsilon}
\def\la{\lambda}	\def\La{\Lambda}
\def\de{\delta}		\def\De{\Delta}
\def\om{\omega}		\def\Om{\Omega}
\def\sig{\sigma}	\def\Sig{\Sigma}
\def\vphi{\varphi}
%
%
\def\CA{{\cal A}}	\def\CB{{\cal B}}	\def\CC{{\cal C}}
\def\CD{{\cal D}}	\def\CE{{\cal E}}	\def\CF{{\cal F}}
\def\CG{{\cal G}}	\def\CH{{\cal H}}	\def\CI{{\cal J}}
\def\CJ{{\cal J}}	\def\CK{{\cal K}}	\def\CL{{\cal L}}
\def\CM{{\cal M}}	\def\CN{{\cal N}}	\def\CO{{\cal O}}
\def\CP{{\cal P}}	\def\CQ{{\cal Q}}	\def\CR{{\cal R}}
\def\CS{{\cal S}}	\def\CT{{\cal T}}	\def\CU{{\cal U}}
\def\CV{{\cal V}}	\def\CW{{\cal W}}	\def\CX{{\cal X}}
\def\CY{{\cal Y}}	\def\CZ{{\cal Z}}
%
%
\font\numbers=cmss12
\font\upright=cmu10 scaled\magstep1
\def\stroke{\vrule height8pt width0.4pt depth-0.1pt}
\def\topfleck{\vrule height8pt width0.5pt depth-5.9pt}
\def\botfleck{\vrule height2pt width0.5pt depth0.1pt}
\def\Zmath{\vcenter{\hbox{\numbers\rlap{\rlap{Z}\kern
		0.8pt\topfleck}\kern
		2.2pt \rlap Z\kern 6pt\botfleck\kern 1pt}}}
\def\Qmath{\vcenter{\hbox{\upright\rlap{\rlap{Q}\kern
                   3.8pt\stroke}\phantom{Q}}}}
\def\Nmath{\vcenter{\hbox{\upright\rlap{I}\kern 1.7pt N}}}
\def\Cmath{\vcenter{\hbox{\upright\rlap{\rlap{C}\kern
                   3.8pt\stroke}\phantom{C}}}}
\def\Rmath{\vcenter{\hbox{\upright\rlap{I}\kern 1.7pt R}}}
\def\Z{\ifmmode\Zmath\else$\Zmath$\fi}
\def\Q{\ifmmode\Qmath\else$\Qmath$\fi}
\def\N{\ifmmode\Nmath\else$\Nmath$\fi}
\def\C{\ifmmode\Cmath\else$\Cmath$\fi}
\def\R{\ifmmode\Rmath\else$\Rmath$\fi}
\def\cadremath#1{\vbox{\hrule\hbox{\vrule\kern8pt\vbox{\kern8pt
			\hbox{$\displaystyle #1$}\kern8pt} 
			\kern8pt\vrule}\hrule}}
\def\proof{\noindent {\underline {Proof}.} }
\def\cqfd{ {\hfill{$\Box$}} }
\def\square{ {\hfill \vrule height6pt width6pt depth1pt} } 
%
%
\def\debut{ \begin{eqnarray} }
\def\fin{ \end{eqnarray} }
\def\non{ \nonumber }
%

%
%
\rightline{SPhT-99-039}
  ~\vskip 1cm
\centerline{\LARGE Influence of Friction on the Direct Cascade}
\bigskip
\centerline{\LARGE of the 2d Forced Turbulence.}
\bigskip
\vskip 1cm
\vskip1cm
\centerline{\large  Denis Bernard
\footnote[1]{Member of the CNRS; dbernard@spht.saclay.cea.fr} }
\centerline{Service de Physique Th\'eorique de Saclay
\footnote[2]{\it Laboratoire de la Direction des Sciences de la
Mati\`ere du Commisariat \`a l'Energie Atomique.}}
\centerline{F-91191, Gif-sur-Yvette, France.}
\vskip2cm
Abstract.\\
We discuss two possible scenario for the direct cascade in two 
dimensional turbulent systems in presence of friction which 
differ by the presence or not of enstrophy dissipation in the 
inviscid limit.They are distinguished by the existence or not 
of a constant enstrophy transfer and by the presence of leading 
anomalous scaling in the velocity three point functions.  
We also point out that the velocity statistics become gaussian 
in the approximation consisting in neglecting odd order 
correlations in front even order ones.

\vfill
\newpage
%

%
%
%

Two-dimensional turbulence has very peculiar properties
compared to 3d turbulence, see eg. ref.\cite{frisch,review} for references.
As first pointed out by Kraichnan in a remarkable paper \cite{kraich},
these open the possibility for quite different scenario 
for the behaviors of turbulent flows in 2d and 3d: if energy and enstrophy
density are injected at a scale $L_i$, with respective rate $\bar \ep$ and 
$\bar \ep_w\simeq \bar \ep L_i^{-2}$, the 2d turbulent systems should
react such that the energy flows toward the large scales and the enstrophy
towards the small scales.  
One usually refers to the infrared energy flow as the inverse 
cascade and to the ultraviolet enstrophy flow as the direct cascade. 
In the (IR) inverse cascade, scaling arguments lead to Kolmogorov's spectrum,
with $E(k)\sim \bar \ep^{2/3}\,k^{-5/3}$ for the energy and 
$(\de u) \sim (\bar \ep r)^{1/3}$
for the variation of the velocity on scale $r$.
In the (UV) direct cascade, scaling arguments give Kraichnan's spectrum
with $E(k)\sim \bar \ep_w^{2/3}\, k^{-3}$ for the energy and 
$(\de u) \sim (\bar \ep_w r^3)^{1/3}$ for the velocity variation.
The aim of this Letter is to discuss a few possible scenario for the
influence of friction on the direct enstrophy cascade of 2d turbulence.
The picture which emerges is that at small scales the velocity field
has a smooth gaussian component, with amplitude diverging in
the frictionless limit, supplemented by possible anomalous corrections
whose characters depend whether there is entrophy dissipation or not 
in the inviscid limit. We also point out that friction provides a way
to regularize amplitudes in the inviscid limit.
\medskip

{\bf A few experimental facts.}
Two-dimensional turbulence has recently been observed in remarkable experiments
providing new informations on these peculiar characteristics.
See refs.\cite{tabel,tabis} for a precise description of 
the experimental set-up. The turbulent flow takes place in a
square cell such that the injection length is $L_i\simeq 10\ cm$.
The bottom of the cell induces friction with a coefficient
$\tau \simeq 25\ s$. The dissipation length $l_d$ is
of order $1\ mm$ and the UV friction length $l_f$ of order
$0.5\ cm$. ($\tau,\ l_d,\ l_f$ are defined below).
The enstrophy transfer rate determined from the three point
structure function has been evaluated in \cite{tabis} 
as $\eta_w\simeq 0.4\ s^{-3}$.
The direct cascade is observed at intermediate scales
between $1\ cm$ and $10\ cm$. Besides the absence of any
experimentally significant deviations from Kraichnan's
scaling, two remarkable facts have been observed:
(i) the vorticity structure functions are almost 
constant on the inertial range with the two-point 
function of order $10\ s^{-2}$;
and (ii) the vorticity probability distribution function
is almost symmetric and almost, but not quite, gaussian. 
\medskip
  
{\bf A model system.}
As usual, to statistically model turbulent flows we consider 
the Navier-Stokes equation with a forcing term.
Let $u^j(x,t)$ be the velocity field for an incompressible fluid, $\nabla\cdot u=0$.
The Navier-Stockes equation with friction reads:
\debut
\d_t u^j + (u\cdot\nabla) u^j - \nu \nabla^2 u^j  + \inv{\tau}\, u^j= -\nabla^jp + f^j
\label{nseq}
\fin
with $p$ the pressure and $f(x,t)$ the external force such 
that $\nabla\cdot f=0$. The pressure is linked to the velocity
by the incompressibility condition: $\nabla^2 p = -\nabla^i\nabla^j u^iu^j$.
The friction term $\inv{\tau}\, u^j$ is introduced in order
to mimick that in physical systems the infrared energy cascade  
terminates at the largest possible scale.
The vorticity $\om$, with $\om=\ep_{ij}\d_iu_j$, is transported by the fluid
and satisfies:
$\d_t\om +(u\cdot\nabla)\om - \nu \nabla^2 \om +\om/\tau= F$ 
with $F=\ep_{ij}\d_if_j$ \cite{cherk}. 
We choose the force to be gaussian, white-noice in time, 
with zero mean and two point function:
$\vev{f^j(x,t)\ f^k(y,s)} = C^{jk}(x-y)\ \de(t-s)$ 
where $C^{jk}(x)$, with $\nabla^j\, C^{jk}(x)=0$, 
is a smooth function varying on a scale $L_i$ and
fastly decreasing at infinity.
The correlation function of the vorticity forcing term is:
$\vev{F(x,t)\ F(y,s)} = G(x-y)\ \de(t-s) $ with  $G=-\nabla^2 \hat C$. 
We shall assume translation, rotation and parity invariance
and expand $C^{jk}(x)$ as: $C^{jk}(x)= \bar \ep\, \de^{ij}
-  \bar \ep_w\, (3r^2 \de^{ij} - 2 x^ix^j)/8+ \cdots$ 
 with $r^2=x^kx_k$. 
The scale $L_i$ represents the injection length. 
The inverse cascade takes place at distances 
$L_i\ll x \ll L_f\simeq \tau^{3/2}\, \bar \ep^{1/2}$.
At finite viscocity, there are two ultraviolet characteristic
lengths, the usual dissipative length $l_d \simeq \nu^{1/2}\, \bar \ep_w^{-1/6}$
and another friction length $l_f \simeq \nu^{1/2}\, \tau^{1/2}$
above which friction dominates over dissipation.
The direct cascade takes place at scale $l_f\ll x \ll L_i$.
We shall consider the inviscid limit at fixed friction,
ie. $\nu\to 0$ at $\tau$ fixed. 
Note that  $(l_f/\, l_d)^2 =\tau\, \bar \ep_w^{1/3}$ is
large in the quoted experiments.

The fundamental property of two dimensional turbulence recognized 
by Kraichnan  \cite{kraich} and Batchelor \cite{batch}
is that the energy cascades towards the large scales because it
cannot be dissipated at small scales. In absence of friction 
the system does not reach a stationnary state. 
But friction provides a way for the energy to escape and
for the system to reach a stationnary state. 
The fact that the energy is not dissipated at small scales translates
into the vanishing of the averaged energy dissipation rate, ie.
$\nu  \vev{(\nabla u)^2} = 0$ in the inviscid limit,
which means that there are no energy dissipative anomaly \cite{batch}, ie.
\debut
\lim_{\nu\to 0} 
\nu  (\nabla u)_{(x)}^2  = 0 \label{noano}
\fin
inside any correlation functions.
However there could be enstrophy dissipative anomalies 
in the sense that $\nu (\nabla \om)^2$ does no vanish in
the inviscid correlation functions:
\debut
\hat \ep_w(x) \equiv \lim_{\nu\to 0}  
\nu  (\nabla \om)^2_{(x)} \label{ano}
\fin
Below we shall discuss possible consequences of
the vanishing or not of $\hat \ep_w$
in presence of friction.
\medskip

{\bf Two and three point velocity correlation functions.}
In absence of energy dissipative anomaly, the mean energy density
relaxes in the inviscid limit  according to
$\d_t \vev{\frac{u^2}{2} } + \inv{\tau} \vev{u^2 } = \bar \ep $,
showing that $\bar \ep$ is effectively the energy injection rate.
It reaches a stationary limit with $\vev{u^2 }=\bar \ep\,\tau$.
Similarly the enstrophy density evolves  according to 
$ \d_t \vev{\om^2} + 2\nu\vev{(\nabla\om)^2} + \frac{2}{\tau}\vev{\om^2}
=2\bar \ep_w$ showing that $\bar \ep_w$ is the enstrophy injection rate. 
Let $\hat \ep_w = \lim_{\nu\to 0} \nu\vev{(\nabla \om)^2}$ be
the enstrophy dissipation rate, $\hat \ep_w< \bar \ep_w$.
In the stationary limit at finite friction
the enstrophy density is finite and equal to:
\debut
\vev{\om^2} = \tau \({ \bar \ep_w - \hat \ep_w }\)
\label{ombar}
\fin
Ie. the enstrophy density is equal to the
difference of the enstrophy injection and the enstrophy dissipation
rates times the friction relaxation time. 
The vorticity two point correlation function
$\vev{\om(x)\om(0)}$ stay finite since it is bounded by $\vev{\om^2}$.
Hence the vorticity correlation cannot diverge as $x\to 0$,
and it cannot have a negative anomalous dimension.
Scalings $\vev{\om(x)\om(0)}\sim r^{-\xi'}$
or $\sim (\log r)^{\xi'}$ with $\xi'>0$ are forbidden in presence of friction.
At short distance one then may have:
\debut
\vev{\om(x)\om(0)}_{\nu=0}
= \bar \Om  - (\tau \bar \ep_w) A\, (r/L)^{2\xi_2} + \cdots \ \label{omega}
\fin
with $\xi_2>0$.
We shall denote the limiting value by $\bar \Om = \tau(\bar \ep_w - \eta_w)$. 
By eq({\ref{3point}) below,  
$\eta_w$ will be identified as the enstrophy transfer rate. 
The amplitude $A$, $\eta_w$ and $\hat \ep_w$, as well as the anomalous
exponant $\xi_2$, are functions of the 
dimensionless parameter $\tau^3\bar \ep_w$.

Since $\nabla^2_x \vev{(\de u)^2}=2 \vev{\om(x)\om(0)}$,
finiteness of the vorticity two point function at coincidant points imply,
$\vev{(\de u)^2)} \simeq \frac{\bar \Om}{2}\, r^2$,
but the subleading terms could have anomalous scalings:
\debut
\vev{(\de u)^2}= \frac{\bar \Om}{2}\, r^2 - 
(\tau\bar \ep_w )\frac{A}{2(\xi_2+1)^2} r^2\, (r/L)^{2\xi_2}
+ \cdots \label{2points}
\fin
The leading scaling is the normal Kraichnan scaling
 but the amplitude is different, 
ie. $\bar \Om$ instead of $\bar \ep_w^{2/3}$.
There is a crossover at scale $r_c$, with $r_c^{2\xi_2}\sim(\bar \Om/\tau\bar \ep_w)$,
above which the second anomalous contribution dominates.
It is interesting to compare the occurence of both 
the smooth $r^2$ term in eq.(\ref{2points}) and the anomalous contribution 
$r^{2\xi_2}$  in eq.(\ref{omega}) with the passive scalar problem 
with friction \cite{cherk}
in which anomalous zero modes occur in the passive scalar correlations
only if the velocity flow is regular.
The smooth $r^2$ term in $\vev{(\de u)^2}$
does not contribute to the asymptotic large $k$ behavior
of the energy spectrum which thus scales as $k^{-3-2\xi_2}$. Namely:
\debut
E(k) \simeq -\frac{2^{2\xi_2+1}\Ga(1+\xi_2)}{\Ga(-\xi_2)}\,
(\tau \bar \ep_w)A\, k^{-3}\, (kL)^{-2\xi_2}
\label{spectrE}
\fin
as $k\to \infty$.
The exponant $\xi_2$ is expected to be universal but not the amplitude $A$.

As usual, stationarity of the two point correlation functions
gives in the inviscid limit, for $x\not= 0$:
\debut
\nabla_x^k \vev{ (\de u^k)\, (\de u)^2 }
+2\vev{(\de u)^2}/\tau &=& 2 (2\bar \ep - \hat C(x) )
\simeq \bar \ep_w\, r^2 + \cdots
\label{dissip} 
\fin
It allows us to determine exactly the three point velocity 
inviscid structure functions:
\debut
\vev{(\de u)^3_{\|}}&=& \frac{\eta_w}{8}r^3 + 
\frac{3A\bar \ep_w}{4(\xi_2+1)^2(\xi_2+2)(\xi_2+3)} 
\, r^{3+2\xi_2} +\cdots \non\\
\vev{(\de u)_{\|}(\de u)_{\bot}^2}&=& \frac{\eta_w}{8}r^3 + 
\frac{3(2\xi_2+3)A\bar \ep_w}{4(\xi_2+1)^2(\xi_2+2)(\xi_2+3)} 
\, r^{3+2\xi_2} +\cdots
\label{3point}
\fin

{\bf Without dissipative anomaly at finite friction.}
If there is no enstrophy dissipative anomalies in presence of
friction, $\hat \ep_w=0$ and hence $\eta_w=0$ assuming continuity
of the correlations. The
three point velocity structure (\ref{3point}) has an anomalous
scaling at leading order. It behaves as 
$$\vev{(\de u)^3} \sim r^{3+2\xi_2}$$ 
since the first term $\eta_wr^3$ vanishes. 
As a consequence the enstrophy transfer is
not constant through the scales, ie. there is no enstrophy
cascade. The anomalous scaling manifests itself only at a subleading 
order in the velocity two point function but the enstrophy
structure function is anomalous:
$$
\vev{(\de \om)^2} = 2 (\tau \bar \ep_w)A(r/L)^{2\xi_2}
$$
Moreover in absence of enstrophy dissipative anomaly the one point
vorticity correlations are gaussian,
\debut
\vev{\exp(s\om)} = \exp(s^2(\tau \bar \ep_w)/2) 
\label{omgauss}
\fin
This is a direct consequence of the stationarity of the vorticity correlations.
It is not true for the higher point functions.
It could  provide a way to check whether the enstrophy
dissipation vanishes or not in presence of friction.
Below we shall present a perturbative argument in favor
of the absence of enstrophy dissipative anomalies in 
presence of friction.

In this scenario with $\hat \ep_w=0$, the quasi symmetry of the
probability distribution may be explained by the fact that
the even order correlations are parametrically larger that 
the odd order correlations, as it is for the two and
three point function, cf eqs.(\ref{2points},\ref{3point}) with $\eta_w=0$.
Namely the even order correlation have leading normal but
anomalous subleading contributions whereas the odd order
correlations would have anomalous contributions at leading order
which decrease faster at small scales.
\medskip

{\bf A scenario for the frictionless limit.}
Experimentally, $(\tau^3\bar \ep_w)$ is large and we are thus
interested in the frictionless limit. We shall discuss 
this limit assuming the absence of dissipative anomaly at $\tau$ finite.
Demanding that the three point structure function
reproduces the known frictionless exact result
with $\vev{(\de u)^3} \sim r^3$ as $\tau\to\infty$ imposes 
that the anomalous exponant $\xi_2$ vanishes and that the amplitude
$A$ goes to one as $\tau \to \infty$.
Moreover, demanding that the velocity structure function is
finite as $\tau \to \infty$ requires:
\debut
\xi_2 = \zeta_2\, (\tau^3\bar \ep_w)^{-1/3} + o((\tau^3\bar \ep_w)^{-1/3})
\quad {\rm and}\quad
A=1 + c\,(\tau^3\bar \ep_w)^{-1/3} +  o((\tau^3\bar \ep_w)^{-1/3})
\label{taularge}
\fin
with $\zeta_2$ and $c$ constant. Since $\xi_2$ is supposed to be universal but
not the amplitude $A$, we expect $\zeta_2$
universal but not $c$.
As a consequence, at fixed non zero distance $r$, the vorticity
correlation behaves logarithmically as:
\debut
\vev{\om(x)\om(0)}\simeq \bar \ep_w^{2/3}\, 
\[\, c - 2 \zeta_2 \log(r/L)\,\] + o((\tau^3\bar \ep_w)^{-1/3})
\label{omlarge}
\fin
Here we assume that the terms represented by the dots in eq.(\ref{omega})
which are subleading, and thus negligeable, at finite friction do
not become predominant in the frictionless limit.
But this is the simplest scenario.

The vorticity structure function, estimated as
$\vev{(\de \om)^2} = 2\tau \bar \ep_w -2 \bar \ep_w^{2/3}
\,( c - 2 \zeta_2 \log(r/L))$,
is then dominated by the constant term since the enstrophy
density diverges as $\tau \to \infty$. This may explain the experimental fact 
that vorticity correlations are almost constant in the inertial range.
This scenario is compatible with experimental data since
experimentally $ \vev{(\de \om)^2}\simeq 10\, s^{-2}$
whereas $2\tau \bar \ep_w \simeq 20\, s^{-2}$ with an estimated error 
of $20\,{}^0/_0$ to $30\,{}^0/_0$.
The velocity structure function will then be:
$$\vev{(\de u)^2} \simeq \bar \ep_w^{2/3}\,r^2\,
\[ c/2 + \zeta_2 - \zeta_2 \log(r/L) + \cdots\]
$$
The smooth $r^2$ is not universal but the logarithmic
correction is (since $\zeta_2$ is expected to be universal).
In the same way the energy spectrum (\ref{spectrE})
$E(k)$ becomes proportional to $k^{-3}$ with:
\debut
E(k)\simeq 2\zeta_2\, \bar \ep_w^{2/3}\, k^{-3} \label{krachn}
\fin
with Kraichnain's constant $C_K=2\zeta_2$ which is universal 
if the anomalous dimensions at $\tau$ finite are universal.
Note that the smooth but non universal $r^2$ terms in the 
velocity correlation does not contribute to the
energy spectrum at large momenta.
Of course, there could be logarithmic correction to the spectrum
if $\zeta_2=0$.
\medskip

{\bf With dissipative anomaly at finite friction.}
If the enstrophy dissipation is non zero in presence of
friction, the enstrophy density is finite but 
smaller than the injected enstrophy density:
$\vev{\om^2} < \tau \bar \ep_w$. The three point
velocity structure function then scales as $\vev{(\de u)^3}
\sim \eta_w r^3$ which allows us to identify $\eta_w$ as
the enstrophy transfer rate, ie. there is enstrophy cascade
with transfer rate $\eta_w$.
The quasi symmetry of the probability distribution function 
would hold only if the entrophy transfer rate is much smaller that the
entrophy injection rate $\eta_w\ll \bar \ep_w$.
There are many scenario for the frictionless limit
depending on how $\eta_w/\bar \ep_w$, which is a function
of $\tau^3\bar \ep_w$, behaves as $\tau\to \infty$.
\medskip

{\bf Neglecting odd order correlations.}
We now discuss what would be the consequences of 
neglecting the odd order correlations in front of the even order ones.
This is suggested by the quasi symmetry of the experimentally
measured probability distribution.
Consider the equations encoding the stationarity 
of an even number of velocity differences, ie.
$\d_t\vev{(\de u^{i_1})\cdots (\de u^{i_{2n}})}=0$. 
Both the terms arising from the advection $(u\cdot\nabla) u$
or from the pressure $\nabla p$ involve correlation
functions with an odd number of velocity insertions.
Assuming that the odd order
velocity correlations are much smaller that the even order ones,
these terms are much smaller and  negligeable 
compared to the remaining terms which are those
arising from the friction and from the forcing and which
involve an even number of velocity insertions. Hence at the leading
order the stationarity condition of the even order correlations
reads:
\debut
\frac{2n}{\tau} \vev{(\de u^{i_1})\cdots (\de u^{i_{2n}})}^{(0)}=
2 \sum_{p<q} (\de C^{i_pi_q}_{(x)})\, 
\vev{\cdots \hat {(\de u^{i_p})}\cdots \hat {(\de u^{i_{q}})} \cdots}^{(0)}
\label{statn}
\fin
with $\de C^{ij}(x) = C^{ij}(0)-C^{ij}(x)$.
The overhatted quantities are omitted in the correlation functions.
In eqs.(\ref{statn}) we have used the fact that 
there is no energy dissipative anomalies. 
The solution of eqs.(\ref{statn}) is provided by a gaussian statistics 
with zero mean and two-point function 
$\vev{ (\de u^i)(\de u^j) }^{(0)} \simeq
\frac{\tau\, \bar \ep_w}{8}( 3r^2 \de^{ij} - 2 x^ix^j)$. 
Hence, in this approximation the velocity structure functions scale as:
\debut
F^{(0)}_{2n}\equiv \vev{ (\de u)^{2n} }^{(0)} &\simeq& 
{\rm const.}\  (\tau \bar \ep_w r^2)^n \label{ansatz}
\fin
This is compatible with eq.(\ref{2points}) since $\bar \Om=\tau \bar \ep_w$
if $\eta_w \ll \bar \ep_w$.
To be consistent it also has to solve the stationarity equations for
the odd order correlations which at this order read:
\debut
\nabla_{x_1}^k \vev{ u^k_{(x_1)}u^{i_1}_{(x_1)}\, u^{i_2}_{(x_2)} 
\cdots u^{i_{2n-1}}_{(x_{2n-1})} }^{(0)} 
+ \nabla_{x_1}^{i_1} \vev{ p_{(x_1)}\, u^{i_2}_{(x_2)}
\cdots u^{i_{2n-1}}_{(x_{2n-1})}}^{(0)} + {\rm perm.} = O(r^{2n})
\label{press}
\fin
with $\nabla^2 p = - \nabla^j\nabla^k u^ju^k$.
Since the statistics is gaussian it is enough to check them for $n=2$.
The gradiant of the pressure has scaling dimension one.
Then the l.h.s. up to $O(r^4)$ is a rank three symmetric $O(2)$ 
polynomial tensor, 
transverse along all coordinates, invariant by translation and
of scaling dimension three. There is no such tensor and thus the l.h.s.
is $O(r^4)$ for $n=2$. In other words, the $n$-point velocity
correlations computed using the gaussian statistics are zero modes. 

In view of the exact formula eqs.(\ref{2points},\ref{3point}), 
this approximation would be better for the transverse velocity correlations.
Note  also that using this approximation
to compute the one point vorticity functions $\vev{\om^n}$ gives 
the exact result (\ref{omgauss}) if there is no enstrophy dissipative anomaly.
Of course there would  be deviations to this approximation:
$$F_{n} \simeq F^{(0)}_{n} + \de F_{n}$$
with corrections $\de F_{n}$ 
including both contributions with normal scaling 
or with possible anomalous scaling as in eq.(\ref{2points},\ref{3point}).
In the scenario without enstrophy dissipation,
these anomalous contributions will be subleading in the
even order correlation but dominant at small scales in
the odd order correlations. On contrary if the enstrophy
transfer is not vanishing these corrections include 
contributions with normal scaling, since $\de F_2\simeq \eta_w\tau r^2$
and $\de F_3 \simeq \eta_w r^3$ as follows from eqs.(\ref{2points},\ref{3point}).
Stationnarity of the three point functions leads then to an equation for
the four point correlations of the form $\nabla^\pi F_4 +\inv{\tau} F_3=0$
with $\nabla^\pi$ the transverse gradiant. Using  $F_3 \simeq \eta_w r^3$
and naive scaling, ie. omitting possible zero mode contributions,
gives a correction $\de F_4 \simeq \frac{\eta_w}{\tau} r^4$ 
to the first order gaussian contribution $F_4^{(0)}$ which satisfies
$\nabla^\pi F_4^{(0)}=0$. Similarly, stationarity of the four point 
functions now give equations of the form 
$\nabla^\pi F_5 +\inv{\tau} \de F_4 \simeq \eta_w \tau\bar \ep_w r^4$.
For $\tau^3\bar \ep_w \gg 1$ as in \cite{tabis}, one
may neglect $\inv{\tau} \de F_4$ in front of $\eta_w \tau\bar \ep_w r^4$,
so that $\de F_5 \simeq \eta_w \tau \bar \ep_w r^5$.
Recursively, naive dimensional analysis apply to stationarity equations with 
$\tau^3 \bar \ep_w \gg 1$ yield to first order in $\eta_w$:
$\de F_{n+2} \simeq \eta_w\, \tau^{1-n}\, (\tau^3 \bar \ep_w)^k\, r^{n+2}$
with $k$ the integer part of $n/3$. Of course $\de F_{2n}\ll F^{(0)}_{2n}$
are small for $\tau^3 \bar \ep_w \gg 1 \geq (\eta_w/\bar \ep_w)$.
But zero modes of the stationarity equations which have been neglected
in this dimensional analysis may give additional contributions
to $\de F_n$. These zero modes will be the leading
terms for the odd order correlations if the enstrophy dissipation
and thus the enstrophy transfer rate vanish.
\medskip 

{\bf MSR formalism.} We now discuss how these scenario could
fit in a field theory approach to the problem. 
The corrections (\ref{taularge}) of the anomalous exponant as a
function of the friction coefficient have a natural interpretation 
in the MSR formalism \cite{msr}.
The MSR formalism provides a way to compute the inviscid vorticity correlations 
using a path integral with action
\debut
S= i\int dt dx\, \varphi_{(x,t)}(\d_t\om + u\cdot \nabla \om + \inv{\tau}\om)_{(x,t)}
+ \half \int dt dx dy\, \varphi_{(x,t)}G_{(x-y)}\varphi_{(y,t)}
\label{action}
\fin
Note that this action describes the inviscid limit $\nu=0$.

Let us first define $\om(x,t)=\bar \ep_w^{1/3} \tilde \om(t\bar \ep_w^{1/3},x)$ 
and $\varphi(x,t)=\bar \ep_w^{-1/3} \tilde \varphi(t\bar \ep_w^{1/3},x)$.
This corresponds to select configurations for which the
dominant time scales and amplitudes are $\bar \ep_w^{-1/3}$. It allows us
to present the theory as a perturbation of the frictionless action $S_0$:
\debut
S = S_0 + ig^{-1/3} \int dt dx\,\tilde \varphi_{(x,t)}\tilde \om_{(x,t)}
\label{pertfric}
\fin
with $g=\tau^3\bar \ep_w$ as dimensionless expansion parameter. 
This shows that the pertubation around the frictionless theory
is controled by $(\tau^3\bar \ep_w)^{-1/3}$ as in eq.(\ref{omlarge}).
Of course implementing this perturbation remains a challenge
as the unperturbed theory is unkown.

Alternatively let us define
$\om(x,t)=(\tau \bar \ep_w)^{1/2} \om'(t/\tau,x)$ and
$\varphi(x,t)=(\tau \bar \ep_w)^{-1/2} \varphi'(t/\tau,x)$.
This corresponds to select configurations with
time scale $\tau$. 
The theory is then formulated as a perturbation of the gaussian action
by the advection term with $g^{1/2}=(\tau^3\bar \ep_w)^{1/2}$ 
as coupling constant:
\debut
S= S_{{\rm gauss}} + i g^{1/2}\int dt dx\, \varphi'_{(x,t)}
(u'_{(x,t)}\cdot \nabla )\om'_{(x,t)} 
\label{pertadv}
\fin
where $S_{{\rm gauss}}$ is the gaussian action obtained
by neglecting the advection term. This is the statistics we found
by neglecting odd order correlation functions. In the unperturbed theory
$\vev{\om^2}=\tau \bar \ep_w$ and there is no anomaly.
Perturbation theory is then computable. We checked that there
is no occurence of the dissipative anomaly in first order 
(one loop) in the perturbation theory.
This is linked to the fact that the friction regularizes the theory
in the inviscid limit such that there is no divergence in the 
Feymann amplitudes, at least in the first order we checked.
This is expected to be true to all orders.
In other words, there is no pertubative contribution
to the enstrophy transfer rate, ie
$\hat \ep_w=\eta_w\simeq 0$ in perturbation theory.

Non perturbative contributions may arise due to instanton
contributions to the path integral. Let us for exemple define
$\om(x,t)=\tau^{-1} \hat \om(t/\tau,x)$ and
$\varphi(x,t)=(\tau^2 \bar \ep_w)^{-1} \hat\varphi(t/\tau,x)$.
It allows us to write the action as $S=(\tau^3\bar \ep_w)^{-1} \hat S$
with $\hat S$ dimensionless. In the saddle point approximation,
instantons could then potentially give a non perturbative contribution 
to the enstrophy transfer rate with 
$\eta_w \sim ({\rm const.})\tau^{-3} \exp(-{\rm const.}/ (\tau^3\bar \ep_w))$.

%
%

\bigskip

{\bf Acknowledgements:} 
We thank  P. Tabeling for motivating discussions on their experiments 
and M. Bauer and K. Gawedzki for numerous friendly conversations.
This research is supported in part by the CNRS, by the CEA and
the European TMR contract ERBFMRXCT960012.

\end{document}